\documentclass[iop,apj]{emulateapj}

\usepackage{xspace}
\usepackage{url}

\shorttitle{Monoceros Ring in Pan-STARRS}
\shortauthors{Slater et al.}

\slugcomment{}

\begin{document}

\title{The Complex Structure of Stars in the Outer Galactic Disk as revealed by
Pan-STARRS1}

\author{Colin T. Slater\altaffilmark{1},
Eric F. Bell\altaffilmark{1},
Edward F. Schlafly\altaffilmark{2},
Eric Morganson\altaffilmark{3},
Nicolas F. Martin\altaffilmark{4,2},
Hans-Walter Rix\altaffilmark{2},
Jorge Pe\~{n}arrubia\altaffilmark{5},
Edouard J. Bernard\altaffilmark{5},
Annette M. N. Ferguson\altaffilmark{5},
David Martinez-Delgado\altaffilmark{6},
Rosemary F. G. Wyse\altaffilmark{7},
William S. Burgett\altaffilmark{8},
Kenneth C. Chambers\altaffilmark{8},
Peter W. Draper\altaffilmark{9},
Klaus W. Hodapp\altaffilmark{8},
Nicholas Kaiser\altaffilmark{8},
Eugene A. Magnier\altaffilmark{8},
Nigel Metcalfe\altaffilmark{9},
Paul A. Price\altaffilmark{10},
John L. Tonry\altaffilmark{8},
Richard J. Wainscoat\altaffilmark{8},
Christopher Waters\altaffilmark{8}}
\affil{}

\altaffiltext{1}{Department of Astronomy, University of Michigan,
    500 Church St., Ann Arbor, MI 48109, USA
    \href{mailto:ctslater@umich.edu}{ctslater@umich.edu};
    \href{mailto:ericbell@umich.edu}{ericbell@umich.edu}}
\altaffiltext{2}{Max-Planck-Institut f\"{u}r Astronomie, K\"{o}nigstuhl 17,
D-69117 Heidelberg, Germany}
\altaffiltext{3}{Harvard-Smithsonian Center for Astrophysics, 60 Garden Street,
Cambridge, MA 02138, USA}
\altaffiltext{4}{Observatoire astronomique de Strasbourg, Universit\'e de
Strasbourg, CNRS, UMR 7550, 11 rue de l'Universit\'e, F-67000 Strasbourg,
France}
\altaffiltext{5}{Institute for Astronomy, University of Edinburgh, Royal
Observatory, Blackford Hill, Edinburgh EH9 3HJ, UK}
\altaffiltext{6}{Astronomisches Rechen-Institut, Zentrum f\"{u}r Astronomie der
Universit\"{a}t Heidelberg, M\"{o}nchhofstr. 12-14, 69120, Heidelberg, Germany}
\altaffiltext{7}{Department of Physics and Astronomy, Johns Hopkins University,
3400 N Charles Street, Baltimore, MD, 21218, USA}
\altaffiltext{8}{Institute for Astronomy, University of Hawaii at Manoa,
Honolulu, HI 96822, USA}
\altaffiltext{9}{Department of Physics, Durham University, South Road, Durham
DH1 3LE, UK}
\altaffiltext{10}{Department of Astrophysical Sciences, Princeton University,
Princeton, NJ 08544, USA}

\begin{abstract}
We present a panoptic view of the stellar structure in the Galactic disk's
outer reaches commonly known as the Monoceros Ring, based on data from
Pan-STARRS1. These observations clearly show the large extent of the stellar
overdensities on both sides of the Galactic disk, extending between
$b=-25^\circ$ and $b=+35^\circ$ and covering over $130^\circ$ in Galactic
longitude. The structure exhibits a complex morphology with both stream-like
features and a sharp edge to the structure in both the north and the south. We
compare this map to mock observations of two published
simulations aimed at explaining such structures in the outer stellar disk,
one postulating an origin as a tidal
stream and the other demonstrating a scenario where the disk is strongly
distorted by the accretion of a satellite. These morphological comparisons of
simulations can link formation scenarios to observed structures, 
such as demonstrating that the distorted-disk model can produce thin density
features resembling tidal streams. Although neither model produces perfect
agreement with the observations---the tidal stream predicts material at larger
distances which is not detected while in the distorted disk model the midplane
is warped to an excessive degree---future tuning of the models to accommodate
these latest data may yield better agreement.
\end{abstract}


\section{Introduction}

The stellar overdensity usually termed the Monoceros Ring (MRi)
has been studied for over a decade, but remains a 
poorly understood phenomenon in the outer Galactic disk. First identified by
\citet{newberg02} and later shown prominently by \citet{yanny03} and
\citet{belokurov06}, in the Sloan Digital Sky Survey (SDSS) the structure
appears as an overdensity of stars at $\sim 10$ kpc from the Sun, spanning
Galactic latitudes from $b\sim+35^\circ$ to the edge of the SDSS footprint of $b\sim
+20^\circ$ and in Galactic longitude extending between $l=230^\circ$ and $l=160^\circ$.

\begin{figure*}[t]
\epsscale{1.1}
\plotone{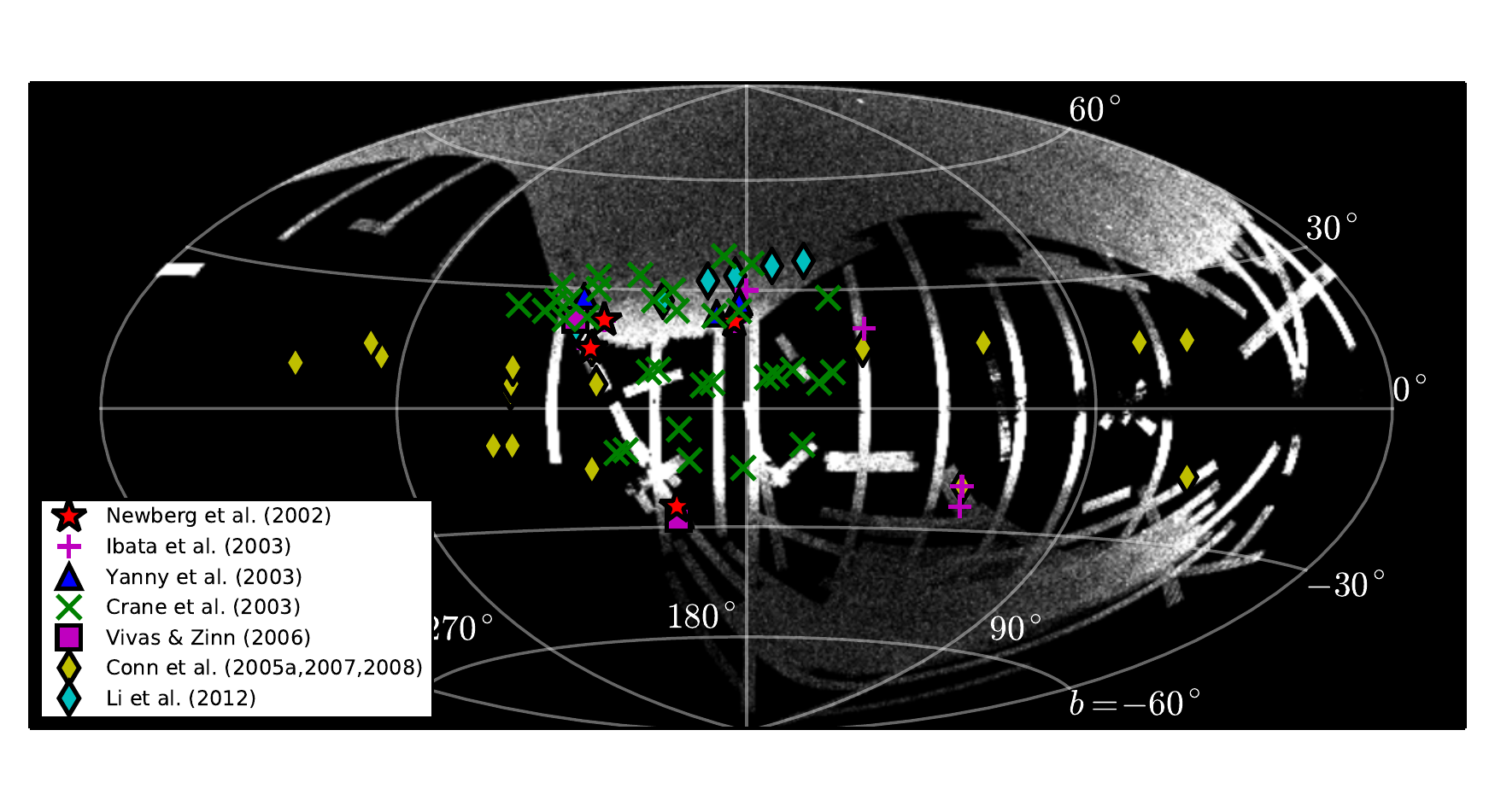}
\caption{A comparison of previous detections of the MRi from various authors
(each listed in the figure legend), overlaid on the map of the MRi as seen by
the SDSS (showing the density of stars with $0.2 < (g-r)_0 < 0.4$ and $18.6 <
g_0 < 19.8$). While the individual pointings clearly show that the MRi occupies
a significant amount of area in the Galactic anticenter, both north and south of
the Galactic plane, it is difficult to understand the morphology of the
structure without contiguous
imaging coverage.
\label{sdss_comparison}}
\end{figure*}

As the initial detections were widely-separated but approximately centered on
the constellation Monoceros, and it appeared to lie at a constant Galactocentric
distance, it was termed the Monoceros Ring \footnote{Though, the structure
clearly extends beyond the borders of the constellation Monoceros, we retain
this terminology for convenience herein.}. Subsequent studies based on modest
numbers of photometric pointings have elucidated the distance dependence of the
structure and provided pencil beam mappings of the structure
\citep[e.g.,][]{ibata03,conn05a,vivas06,conn07,conn08,conn12,li12}. These pointings
have also shown that the feature appears both north and south of the Galactic
plane at similar Galactic longitudes \citep{conn05a,dejong10}, further expanding
the known size of the structure. A summary of many of the detections of the MRi
is shown in Figure~\ref{sdss_comparison}, along with the MSTO stellar density
map from the SDSS showing the MRi detections within its footprint.
Spectroscopic observations have shown that much of the MRi is consistent with a
nearly circular orbit at a velocity of $\sim 220$ km s$^{-1}$ \citep{crane03,
conn05b, martin06} and have potentially identified related star clusters at
similar velocities \citep{frinchaboy04}. The association between the MRi and
other density structures in the Galactic disk and halo has been the source of
considerable controversy, with the Canis Major overdensity
\citep{martin04} and the Triangulum-Andromeda overdensity \citep{rochapinto04}
both lying near detections of the MRi, and with debate as to whether the
density structure seen in SDSS is of a common origin or multiple distinct
structures \citep{grillmair06,grillmair08}.

While the basic observations of the MRi are generally agreed on, there is very
little consensus on details beyond these, and particularly in the origin of the
structure there is wide disagreement. One possibility is that the MRi is the
tidal debris from a disrupting dwarf satellite galaxy
\citep{martin04,penarrubia05,sollima11}. In this scenario the stream's orbital
plane is similar to that of the Galactic disk by virtue of a low-inclination
progenitor orbit. An alternative scenario is that the stars in the MRi
originally formed in the Galactic disk, but were stirred up by some dynamical
perturbation to heights of 1-5 kpc above and below the disk. Qualitatively, such
scenarios can be simply stated, but their parameterization and characterization
can be complex.

Some models have sought to characterize the observations of overdensity by
modeling a flare and a warp in the Galactic stellar disk
\citep{momany06,hammersley11}, while numerical models have sought to recreate a
MRi-like feature in N-body simulations by perturbing a disk with satellite
galaxies \citep{younger08,kazantzidis08}. Perturbations to the disk by
satellites have been studied both in observations \citep{widrow12} and in N-body
simulations of spiral arms \citep{purcell11} and vertical density waves
\citep{gomez13}, all reinforcing the picture that the disk can exhibit complex
structure in response to close satellite passages. Observationally, in M31
disk-like stellar populations have been found at large distances from the bright
stellar disk \citep{richardson08}, and appear to also have disk-like kinematics
\citep{ibata05}. Distortions and warps in the outer stellar disks and HI gas of
galaxies appear common \citep{sanchez90,garciaruiz02}, and some have been shown
to cause the formation of young stars at large heights from the plane as defined
by the central region of the galaxy \citep{radburnsmith14}.

Most of the models of the Monoceros Ring can produce qualitative matches to
the available data, which has left little leverage to distinguish between these
scenarios. In particular, the lack of a contiguous map of the MRi has
forced models to rely on matching the distances and depth of the structure as
seen by the  available sparse pointings. This limited dataset has made
it difficult to see a correspondence (or disagreement) between the N-body
simulations and the actual MRi structure, since the complex density structure
predicted by these simulations can be masked when only a limited number of
discontiguous pointings are available for comparison.

In this work we present a panoramic view of the
Monoceros Ring using Pan-STARRS1 \citep[PS1;][]{kaiser10}, extending both
north and south of the Galactic disk and covering $160^\circ$ in Galactic
longitude. This is the most comprehensive map of the MRi to date, enabled by
extensive sky coverage of PS1 and its survey strategy that --- unlike SDSS ---
fully includes the Galactic plane. As we will show, this panoptic view
provides a dimension of spatial information that had previously only been hinted
at and incomplete, particularly in the southern Galactic hemisphere.

To illustrate the utility of these new maps, we also present a first comparison
to physically motivated N-body simulations. Though qualitative in nature, we
will show that such comparisons can immediately be used to refine our
understanding of the physical ingredients necessary for reproducing the
structure. In the following work we describe the PS1 survey and the data
processing in Section~\ref{sec_data}, followed by a discussion of the resulting
MRi maps in Section~\ref{sec_obs}. We show the comparisons to the two N-body
models in Section~\ref{sec_modeling}, and we discuss the results and conclusions
in Section~\ref{sec_conclusions}.

\section{The Pan-STARRS1 Data}
\label{sec_data}

PS1 is a 1.8m telescope on the summit of Haleakela, Hawaii, which operates as a
dedicated survey instrument (Chambers et al., in preparation). The telescope
images in five bands
($g_\mathrm{P1}r_\mathrm{P1}i_\mathrm{P1}z_\mathrm{P1}y_\mathrm{P1}$) with
average exposure times on the order of 30-45s \citep{metcalfe13}. The individual
exposures are photometered by an automated pipeline \citep{magnier06} and
calibrated to each other self-consistently using partially overlapping exposures
\citep{schlafly12}, yielding a calibration precision better than $10$ mmag as
measured against SDSS. In this work we use data from the $3\pi$ survey, which
covers the entire sky north of declination $-30^\circ$ and is designed to obtain
approximately four exposures per pointing, per filter, per year. Although the
survey does produce stacked images and the resulting photometry, in this work we
use data that is the merger of the photometric catalogs of all individual
exposures. Stacked data from PanSTARRS-1 will reach more than a magnitude
deeper, but the processing pipeline for the single epoch data is currently more
mature and reliable. MSTO stars in Monoceros are easily detected in the PS1
single epoch images, and so we accordingly use that data in this work.  We use
the most recent consistent reprocessing of the $3\pi$ observations, termed
Processing Version 1 (PV1), which contains observations obtained primarily
between May 2010 and March 2013. As measured against the SDSS stripe 82 coadd
catalog, our 50\% completeness levels range from roughly $g_{P1} =21.4$ to
$22.0$, and $r_{P1} = 21.2$ to $21.8$ \citep{slater13}. Since these are limits
are substantially fainter than our target MSTO stars in the MRi, this limited
photometric depth will not impair our results.

Because the stars used we focus on in this work are much brighter than the
completeness limit, our photometric uncertainties are largely the result of
large scale systematic effects and calibration uncertainties rather than photon
noise on the photometry itself. In general this results in a typical
uncertainity of $0.01-0.02$ magnitudes 
\citep[see][for further details]{schlafly12}.
The most significant remaining uncertainity is the correction for Galactic
extinction. The relatively narrow color selection we use makes the number counts
of MSTO stars sensitive to small color shifts, which can be caused by errors in
the extinction maps used for dereddening. For extinction correction we use the
maps of \citet{SFD}, corrected with the factors prescribed by
\citet{schlafly11}. While the \citet{SFD} maps are in general excellent, there
are regions of our maps where changes in the measured stellar density correlate
strongly with dust extinction features. This is primarily an issue within
$20-30^\circ$ of the north celestial pole, and we therefore do not want to
overinterpret the results there. Beyond this particular region the extinction
correction appears to be well-behaved and there are fewer correlations between
MSTO map features and extinction features. The regions where we have marked
Monoceros-like features do not show significant dust features.

Here our principal objective with the PS1 observations is to create a series of
stellar density maps showing the spatial extent and morphology of the MRi in a
low-latitude version of the ``Field of Streams'' \citep{belokurov06}. To do
this, we impose cuts on color and magnitude of stars in a way designed to select
main sequence turn-off (MSTO) stars of an old ($\sim 9$ Gyr) population with
$-1 \lesssim \rm{[Fe/H]} \lesssim 0$ and at the range of heliocentric distances
of interest. A color range of $0.2 < (g-r)_0 < 0.3$ optimizes the contrast
between the MRi and any foreground (nearby Galactic disk) or background (stellar
halo) contamination. 
We have estimated the distance to the stars selected by this color cut 
with the BaSTI set of isochrones \citep{pietrinferni04}. We use a 9 Gyr old,
[Fe/H]=-1.0 isochrone populuated with a \citet{kroupa01} initial mass function
and with realistic observational uncertainties added. The choice of stellar
population parameters are in line with the metallicity measured for the MRI by
both \citet{conn12} and \cite{meisner12}, though substantial scatter in the
metallicity of the MRi has been reported \citep[e.g.,][]{yanny03,crane03}.
The median magnitude of the synthesized stars selected by this color cut is
$M_{g,P1}=4.4$, which we shall adopt for our quoted distances, though the spread
is considerable and 70\% of synthesized stars are found within $\pm 0.5$
magnitudes of the median value. The uncertainity in the stellar populations of
the MRi, particularly the age, adds an additional $\sim0.2$ magnitude systematic
uncertainity, but this is substantially less than the intrinsic magnitude spread
of the MSTO in a single stellar population. The uncertainity between different
sources of isochrones is at a similar level, and studies of globular clusters
also produce similar results \citep{newby11}.

\section{Observed MRi Morphology}
\label{sec_obs}

\begin{figure*}[t]
\epsscale{1.1}
\plotone{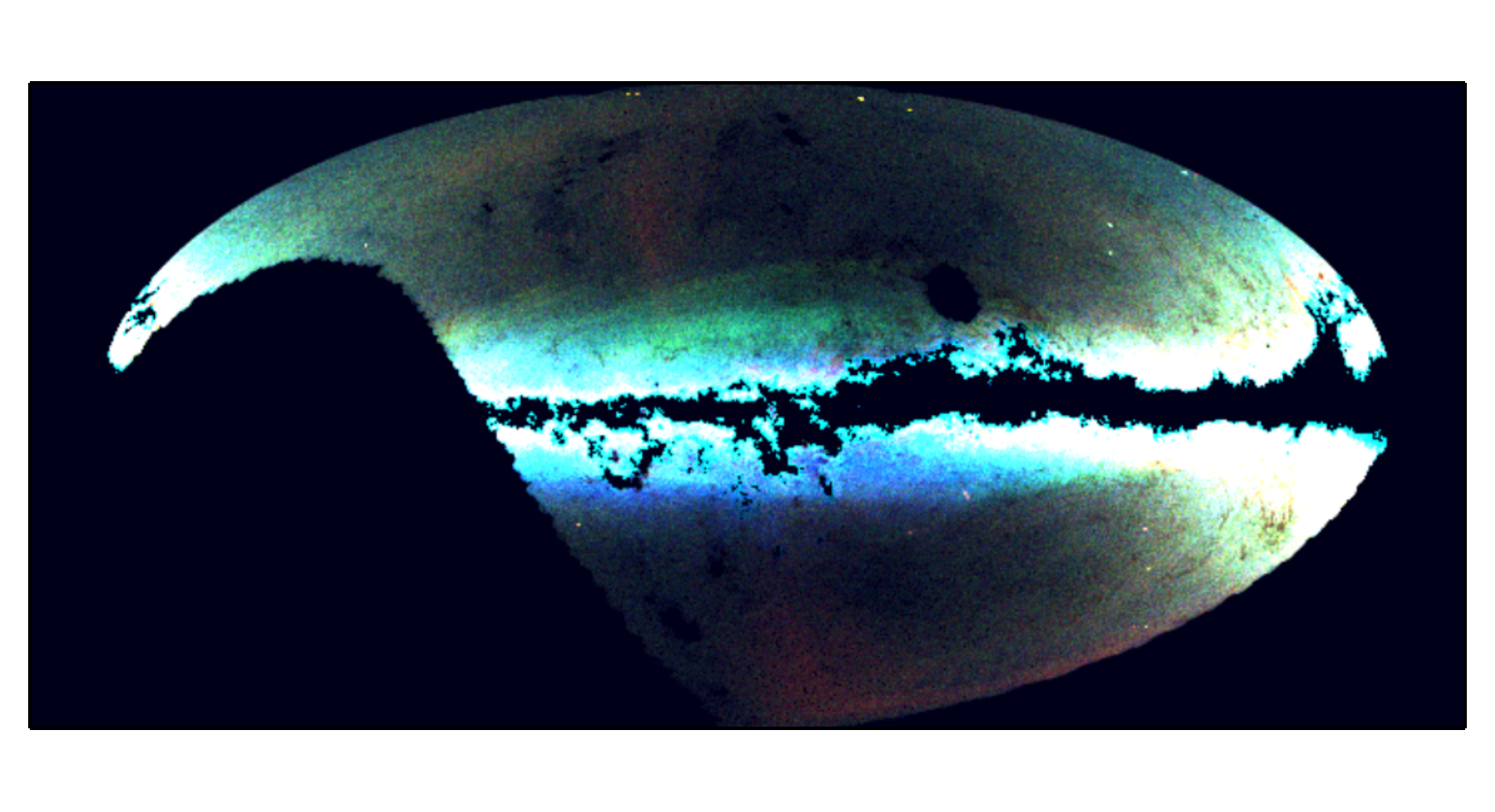}

\caption{Pan-STARRS1 map of star counts in Galactic coordinates for stars with
$0.2 < (g-r)_0 < 0.3$. Nearby stars with $17.8 < g_0 < 18.4$ (4.8-6.3 kpc) are
shown in blue, stars with $18.8 < g_0 < 19.6$ (7.6 - 11.0 kpc) are shown in
green, and more distant stars with $20.2 < g_0 < 20.6$ (14.4 - 17.4 kpc) are
shown in red. The Galactic anticenter is in the middle, and the Galactic center
is on the right edge. The Monoceros Ring can clearly be seen in broadly
horizontal green structure on the nothern side of the plane and in the similar
structure on the southern side of the plane in blue, both of which extend over
$130^\circ$ in Galactic longitude. The difference in color as presented suggests
that the southern component is slightly closer to the Sun than the northern
component. The Galactic plane and some localized regions near the plane are
missing due to high extinction, while the apparent hole near the north celestial
pole was imaged but not included in this processing of the data. There are some
regions of the north Galactic cap and near the celestial pole that suffer from
poor PS1 coverage. The Sagittarius stream appears nearly vertical in red on both
sides of the disk. This figure is available in FITS format in the online
version of the journal.
\label{img_on}}
\end{figure*}

\begin{figure*}[t]
\plotone{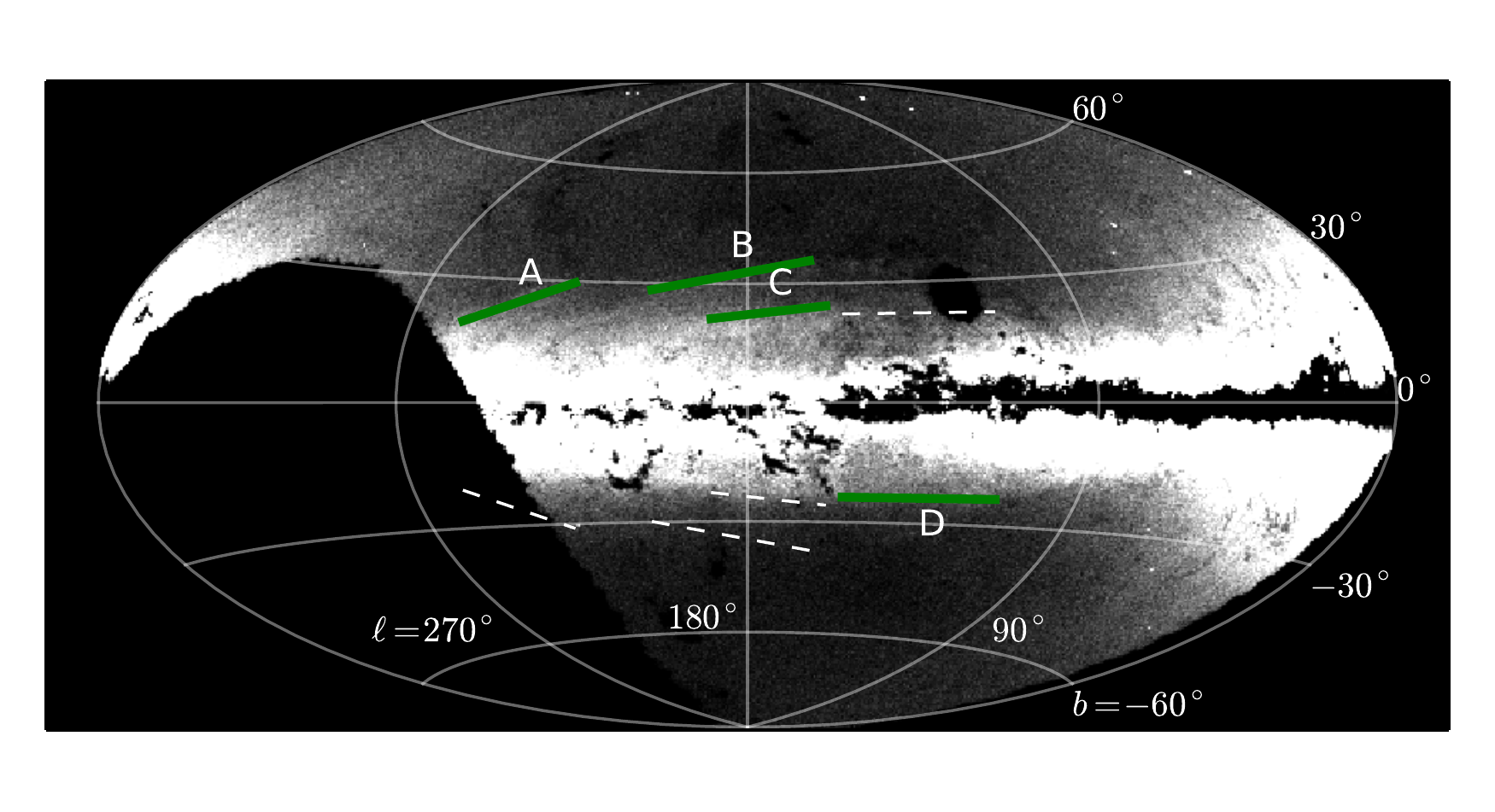}
\caption{Same as Figure~\ref{img_on}, showing the middle (green in
Figure~\ref{img_on}) distance slice
and with several density features labeled.  
These markings are not intended to be comprehensive, and many of the
features extend beyond the extent of the labels. The white dashed lines show
the location of the labeled features reflected across the Galactic plane. The
grid shows $l=90^\circ$ (right side), $l=180^\circ$, and $l=270^\circ$ (left
side), along with lines at $b=\pm 30^\circ$ and $\pm 60^\circ$.
\label{img_labeled}}
\end{figure*}

\begin{figure*}[t]
\epsscale{1.2}
\plotone{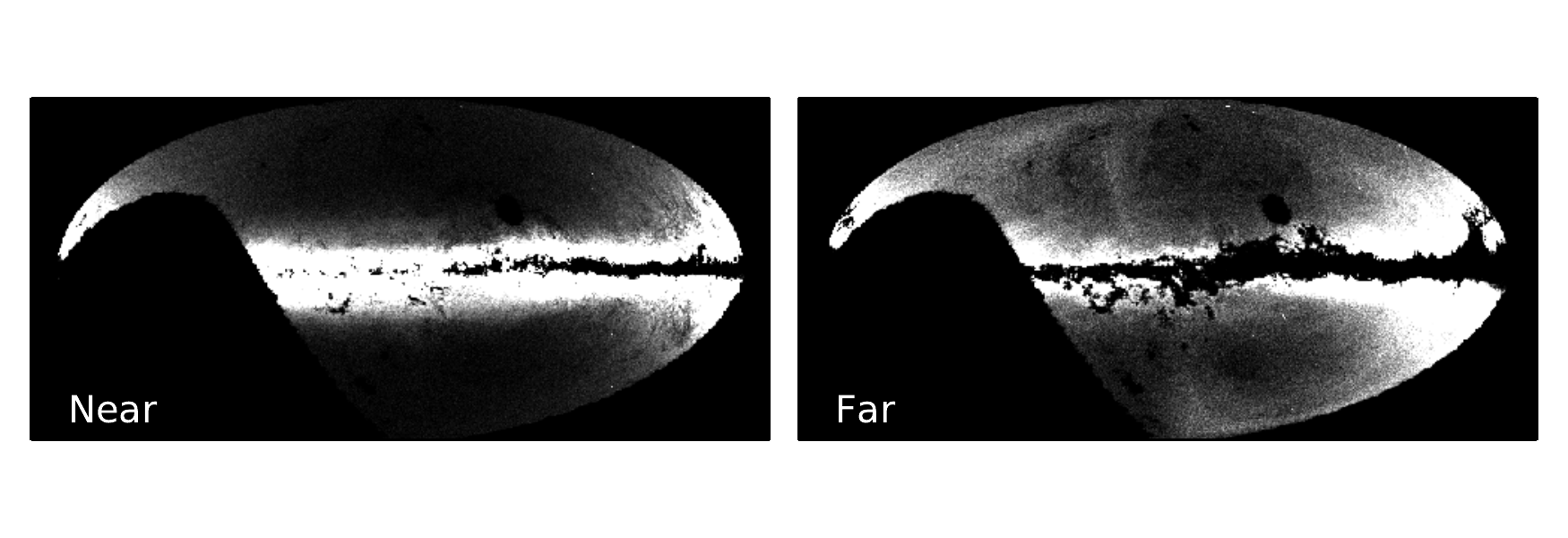}
\caption{Two distance slices, one nearer than the main body of the MRi (left,
shown in blue in Figure~\ref{img_on}, $17.8 < g_0 < 18.4$) and one further
(right, shown in red in Figure~\ref{img_on}, $20.2 < g_0 < 20.6$). In the nearer
slice there is very little evidence of the MRi in the north, though southern
structure remains visible. In the far slice the MRi has become much less
prominent and there is little new MRi-related structure that becomes apparent.
\label{img_slices}}
\end{figure*}

Figure~\ref{img_on} shows the MSTO stellar density map, projected in Galactic
coordinates, centered on the Galactic anticenter. The components of the
three-color image were chosen to show stars with $17.8 < g_0 < 18.4$
(centered on roughly 4.8-6.3 kpc, but also broadened by the intrinsic
MSTO magnitude spread) in blue, stars with $18.8 < g_0 < 19.6$ in green
(7.6-11.0 kpc), and more distant stars with $20.2 < g_0 < 20.6$ in red
(14.4-17.4 kpc). These maps are available as FITS files in the online
edition of the journal, along with the intermediate distance slices. The most
prominent features of the MRi are the broad horizontal arcs on both the northern
and southern sides of the disk, primarily in blue and green, showing several
sharp density features at large heights above the disk. On the northern side
multiple arcs are visible, which we have labeled in Figure~\ref{img_labeled} for
convenience in describing them. The MRi features seen in SDSS are what we have
labeled Features~B and~C, with some small part of Feature~A also visible towards
the edge of the SDSS coverage. Where the Pan-STARRS and the SDSS coverage
overlap there is good agreement on the morphology of the features, while the
additional new area available in the PS1 coverage shows all of these features in
a contiguous map, revealing that they extend substantially beyond the SDSS
footprint.

The southern extent of the MRi has not been seen before in a wide-area map. The
broad southern sharp-edged arc is strikingly similar to the observed arc on the
northern side of the disk, particularly Feature~C, and leaves little doubt that
these features are related. In our maps the MRi clearly encompasses a vast area
of the Galactic anticenter region, spanning from $b=-25^\circ$ to $b=+35^\circ$
and covering nearly $130^\circ$ in longitude on both sides. It is interesting to
note that the material that makes up Features~C and D appears to blend smoothly
in with the disk closer to the Galactic plane, with no second sharp edge at
lower latitudes to denote an ``end'' of the MRi material. This is particularly
apparent in the south, as some extinction features may be affecting the north
slightly more.

Though the bulk of the MRi appears similar on both sides of the Galactic plane,
there are small but noticeable asymmetries between the northern and southern
features. To aid in seeing this, the marked features have also been reflected
across the Galactic equator and denoted with dashed white lines. The A and B
features clearly extend further off the Galactic plane than any feature in the
south, though C and D seem to be very similar in extent both in latitude and
longitude. There does not appear to be the same multiplicity of arcs on the
southern side as compared to the north.

These arcs, which we refer to as Features~A and~B, have been previously pointed
out by \citet{grillmair06} and revisited in \citet{grillmair11}, which referred
to these features the anticenter stream and the eastern banded structure,
respectively. Though these stellar density features certainly exist, 
their decomposition into ``distinct'' features does not appear obvious or
unique.

In Figure~\ref{img_slices} we show the nearer (blue in Figure~\ref{img_on}) and
farther (red in Figure~\ref{img_on}) distance slices separately from
Figure~\ref{img_on}, so that they can be examined independently. These maps show
that the structure is relatively well-confined in heliocentric distance, with
the southern part becoming visible in the near slice and only hints of the
structure remaining in the far slice. We will illustrate the utility of these
distance slices for constraining models in Section~\ref{sec_modeling}, but from
the data alone we can show that the structure is not very extended in
heliocentric radius. There is, however, an offset in distance between the
northern and southern components of the MRi, with the southern component
somewhat closer to the Sun than the northern side.

\section{Model Comparisons}
\label{sec_modeling}

In order to guide the understanding of the observed MRi, we have
created ``mock observations'' of two N-body simulations. One of these
models the MRi as perturbed disk stars that have been stirred up by satellite
galaxies \citep{kazantzidis09}, while the other models the MRi as simply the
debris from a disrupted satellite \citep{penarrubia05}. These two
simulations serve to illustrate the range of morphologies that these
categories of models generate, along with demonstrating the utility of the PS1
maps for differentiating between these models. We note that at this stage
neither simulation has been tuned to reproduce the PS1 observations, so
discrepancies must be expected, but our goal is to highlight these discrepancies
so that future models may be better tuned to match the observations.

To produce maps of the simulations with similar observational characteristics to
the PS1 data, we use the same isochrones as in Section~\ref{sec_data} to
determine the extent to which each simulation particle contributes to the
various magnitude slices, with the considerable magnitude spread of the MSTO
causing simulation particles to potentially contribute to multiple 
slices. As per our population modeling in Section~\ref{sec_data}, we
account for this spread by approximating the MSTO absolute magnitude
distribution as a Gaussian centered on $M_{g,P1}=4.4$ and a width of $0.5$
magnitudes. This center and width approximate the synthesized distribution of
MSTO stars, though in detail the shape likely deviates somewhat from a Gaussian.
After determing each particle's contribution to a given magnitude slice, the
particles are projected onto the ``sky'' as would be seen by an observer and
summed to produce the star counts map. It is important to note that the
simulations are run at resolutions much coarser than single stars, and
hence individual simulation particles are visible in some regions of the maps
which contributes to a ``grainy'' appearance. We do not correct for this. 

\subsection{Satellite Accretion Model}

As an example of models that recreate the MRi as the tidal debris
of a dwarf galaxy, we show the simulations of \citet{penarrubia05}. This
simulation attempted to reproduce all of the positions, distances and
velocities of the MRi that were known at the time of publication.
To do so, the authors varied the properties of the accreted dwarf along with its
orbit and the shape of the Galactic potential to find a solution that best
reproduced the known observations. The resulting best fit has the hypothesized
dwarf on a very low-eccentricity orbit ($e\sim0.1$) and at a low inclination
relative to the Galactic plane, which allows it to make multiple wraps over a
relatively narrow range of Galactocentric radii and relatively close to the disk in
height. The simulation is also designed so that the main body of the disrupting
satellite appears in the region of the Canis Major overdensity
\citep{martin04}, as an attempt to link the two structures. As more recent
evidence suggests that the Canis Major overdensity may not be related to a
disrupted dwarf \citep{mateu09}, in our comparison we will focus on the general
behavior of the tidal stream component rather than the specific location of the
progenitor. Also note that this simulation focused on the properties of tidal
debris, there are no N-body particles from the Galactic disk, which is instead
modeled with a static potential. The particles used to reproduce the stream are
the dark matter particles from the satellite, as there was no distinction made
between a central concentration of luminous particles and a larger dark matter
halo. As a result, the model predictions of the surface brightness of the stream
along its orbit may be inaccurate, and further work would be needed to make
more precise predictions. Our focus is thus on the general morphological
comparison between the model and observations.

\begin{figure*}[t]
\epsscale{1.2}
\plotone{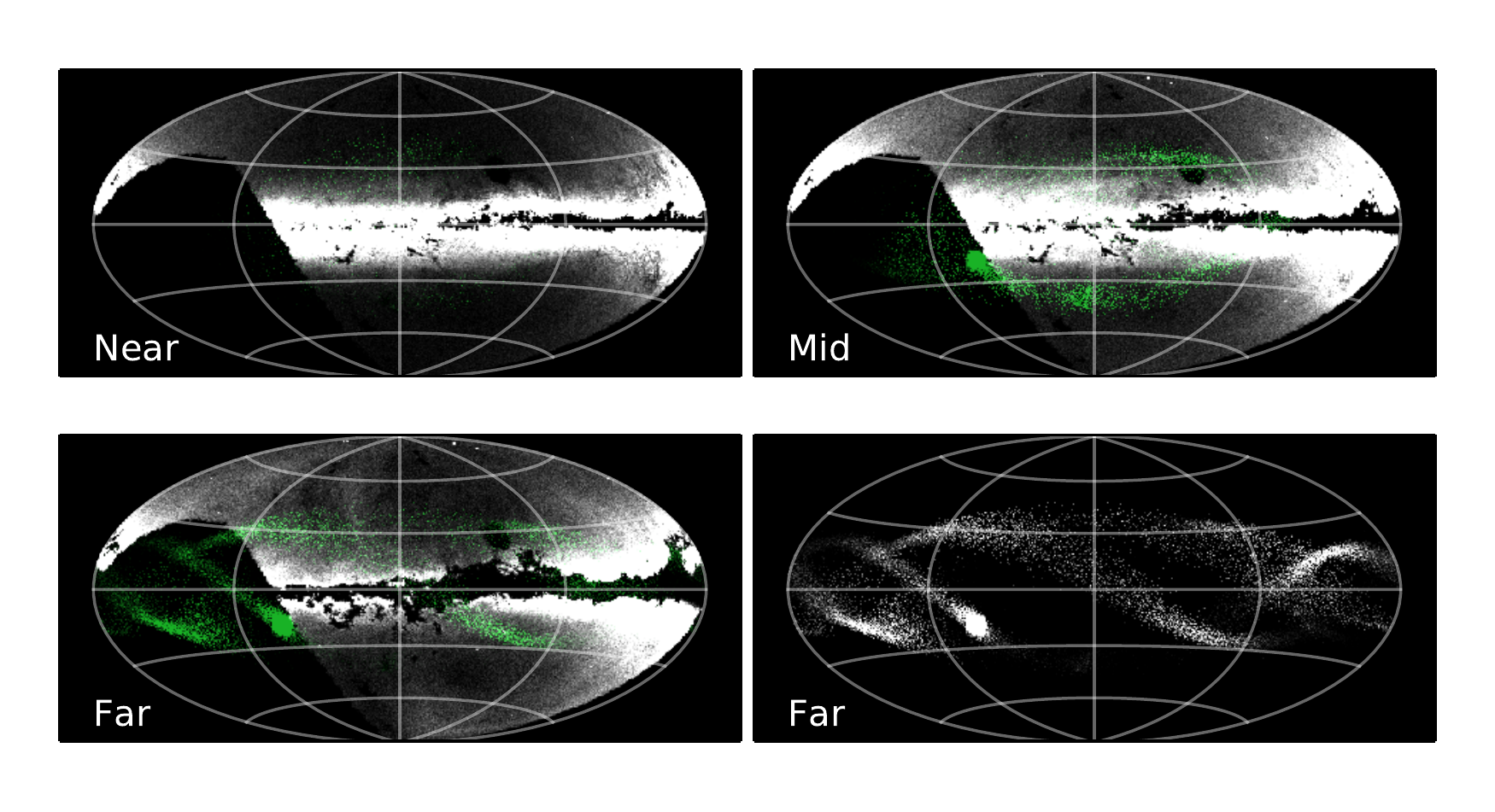}
\caption{Visualization of the \citet{penarrubia05} model for the formation of
the MRi by accreted satellite material (green points), plotted on top of the the
PS1 observations. We show three magnitude slices
(approximately corresponding to distance slices used in the observations), with
the top-right ``mid'' slice matching Figure~\ref{img_on} and the ``near'' (top
left) and ``far'' (bottom left) slices matching Figure~\ref{img_slices}. The
``far'' simulation slice is repeated by itself on the bottom right for clarity.
The PS1 images are the same as Figures~\ref{img_on} and \ref{img_slices}. The
``mid'' slice shows broad agreement with the observed structure, though somewhat
offset, while the ``far'' slice shows a considerably different set of
overdensities that do not appear to match the observations.
\label{jorge_combined}}
\end{figure*}

A visualization of the simulation  can be seen
in Figure~\ref{jorge_combined} as green points (satellite debris particles)
plotted on top of the observed PS1 data.  Three magnitude slices are plotted,
corresponding to the cut targeting the main body of the MRi in
Figure~\ref{img_on} (labeled ``mid'') and the two cuts on the near side and far
side of the MRi from Figure~\ref{img_slices}. From this we can see that the
overall shape of the MRi is reproduced quite well in the ``mid'' distance slice
of the simulation. The north and south both exhibit very broad structures with a
sharp edge on the side away from the Galactic plane, along with a convincing
degree of symmetry across the two hemispheres that is reminiscent of the
symmetry of Features C and D. 

Though the ``mid'' distance slice exhibits considerable resemblance (though with
some spatial offset), there is substantial discrepancy between the simulation
and the observed structure in the ``far'' slice. 
Some of the material in the far slice does trace northern and southern
edges of the MRi as observed, but generally the material appears in new regions
of the sky that were not as well-populated in the ``mid'' slice. This is
noticible in the south, where a broad stream paralelling the observed MRi in the
``mid'' slice is replaced by a different wrap of the stream at a different angle
in the far slice, cutting across the plane rather than paralleling it. In the
north the simulated MRi material is prominent in differing regions; while both
match some portion of the MRi, the mid and far slices do not match each other. These
features are in contrast to the strong similarity between the observed mid and
far slices, in which the far slice does not show any new features of MRi
appearing that were not prominent in the mid slice.
Some of the structures that appear
most prominently at closer distances still appear in the far slice, due to the
intrinsic spread in magnitude of the MSTO, but no new components of the MRi
appear in the far slice. A portion of the features in the simulation do fall
outside the PS1 footprint on the sky, particularly structures on the far left
side of the figure, which corresponds to the southern celestial hemisphere, but
there appears to be substantial structure even within the area covered by PS1.
The recent deep mapping near Andromeda \citep{martin14} may show some material
resembling the stream model shown here (which they refer to as the
``PAndAS MW stream'', near $l=120^\circ$, $b=-20^\circ$), but the detected
material is further than both our ``far'' slice and the predicted stream, so the
correspondence is tentative. The overdensity in the simulation between
$l=180^\circ$ and $l=270^\circ$ and just south of the Galactic plane is the main
body of the disrupting dwarf. We do not find an overdensity of this significance
in the PS1 data, but this is mostly off the edge of the PS1 coverage and is also
likely to be a specific prediction of this particular simulation rather than a
general prediction of tidal stream models. 

The general discrepancy between the simulation and the observed MSTO maps in the
amount of structure at large distances may be intrinsic to satellite accretion
models, which in general require dwarfs to be on at least mildly eccentric
orbits and thus causing the widely spread debris. The challenge for future
accretion models that attempt to reproduce the MRi is thus to plausibly explain
the circular orbit, or to present some other way in which the debris at larger
distances is hidden from view.

\subsection{Disrupted Disk Model}

\begin{figure*}[t]
\epsscale{1.0}
\plotone{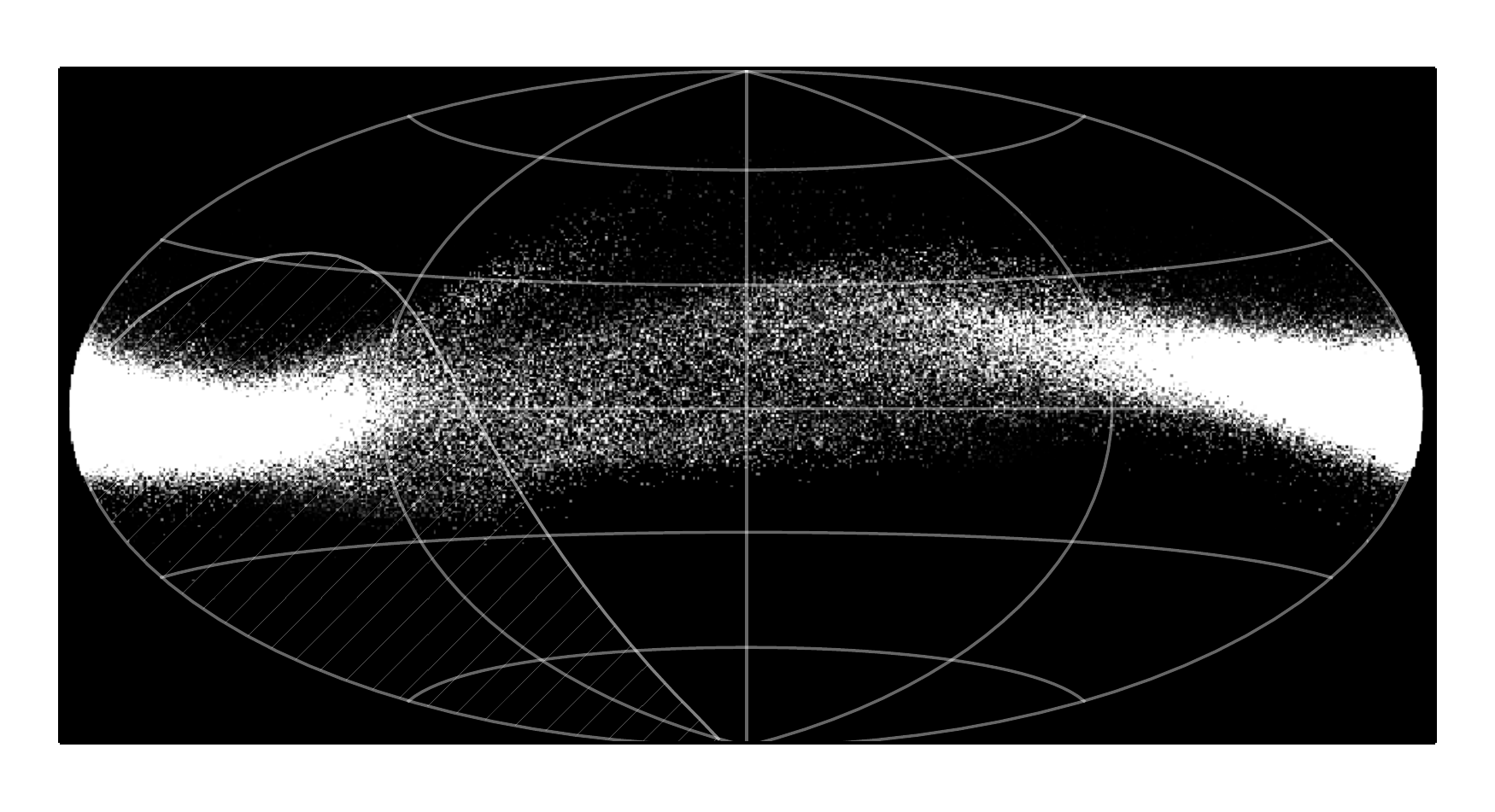}
\caption{Visualization of the \citet{kazantzidis09} model, approximating the
distance cut and uncertainties from the PS1 data in Figure~\ref{img_on}. The
feature annotations from Figure~\ref{img_labeled} have been included to give a
sense of scale of the out of plane features. The hatched region indicates the
area south of declination $-30^\circ$, which is not observed by PS1. For clarity
and as the model already includes a galactic disk, we have not overplotted it on
the existing data. While the simulation is not designed to replicate individual
features, there is a striking similarity between the model and the observed MRi
in the presence of thin wisp-like features, but also a clearly excessive level
of warping of the disk midplane beyond that seen in the Milky Way.
\label{stelios_on}}
\end{figure*}

\begin{figure*}[t]
\plotone{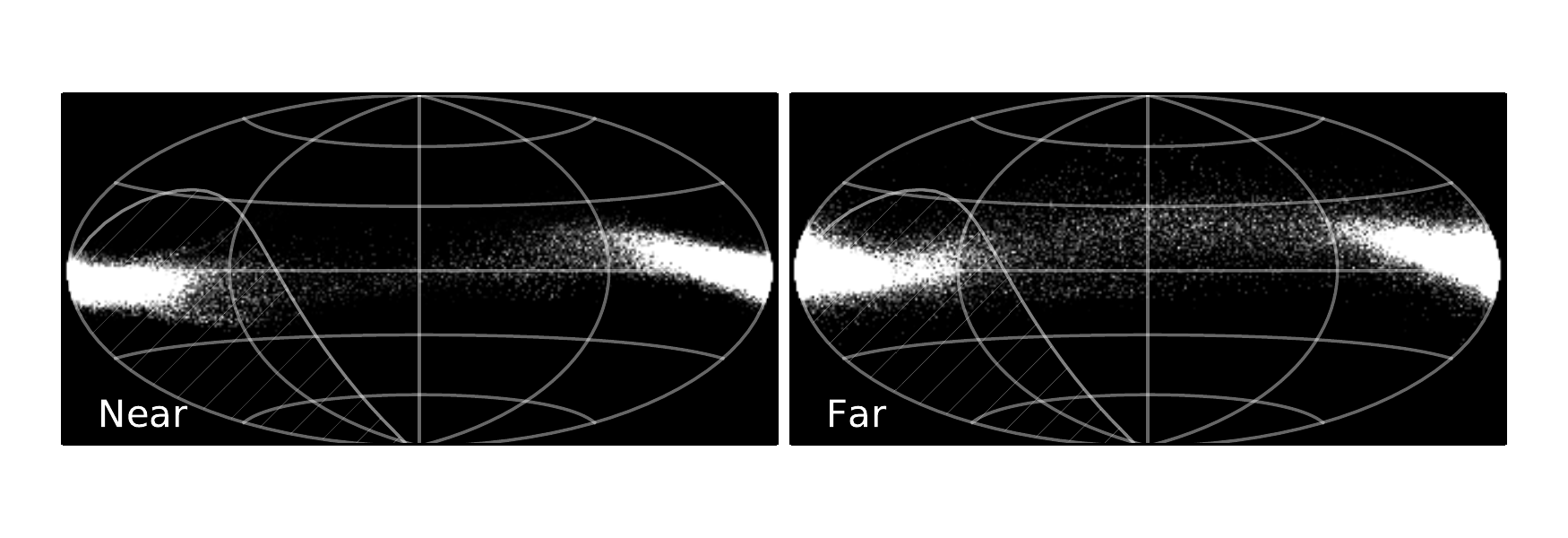}
\caption{Visualization of \citet{kazantzidis09}, showing a closer and a further
distance cut (same as in Figure~\ref{img_slices}). \label{stelios_combined}}
\end{figure*}

The simulation of \citet{kazantzidis09} present a case where the disk of the
galaxy has been strongly disrupted by the impact of satellite galaxies. The
simulation we show is one of a suite of controlled experiments where a Milky
Way-sized disk was subjected to bombardment by a cosmologically-motivated set of
six dark halos. These halos range from 20\%-60\% of the mass of the disk itself,
with the majority of the effect on the disk being driven by the most massive
accretion event. The disk in the simulation has a mass of $3.53 \times 10^{10}$
$M_\odot$, and the satellite halos hence had masses ranging from $7.4 \times
10^9$ $M_\odot$ to $2 \times 10^{10}$ $M_\odot$. The pericenters of the
satellite orbits ranged from 1.5 kpc to 18 kpc. In contrast to the
\citet{penarrubia05} model, in this visualization there are no
particles from the dwarfs shown; the particles we show were all originally in
the simulated disk.

While the transformation from simulation particles into ``mock'' MSTO slices is
the same for this simulation as for our presentation of the \citet{penarrubia05}
simulation, there are a few additional complications.  Because the
\citet{kazantzidis09} simulation was not tuned to reproduce any observations of
the MRi, there is no preferred position for the observer. That is, we can
visualize the simulation as if we were at any point on the Solar circle, each
time obtaining a unique view of the disrupted disk.  We chose an observer
position that gives the best qualitative resemblance between the simulations and
the observations in order to show as many positive features of this type of
model as possible. There is also some question as to how best to define the
Galactic plane in such a simulation. A gas disk would be the natural choice, but
since this is an N-body only simulation, that option is not available. Following
the initial analysis of the simulation in \citet{kazantzidis09}, we have chosen
to align the galactic plane of our visualization perpendicular to the total
angular momentum axis summed over all of the particles in the simulation.

The resulting mock observations are shown in Figure~\ref{stelios_on}, where a
substantial warping of the disk is clearly evident. To the right of the
anticenter there is very little material remaining along the Galactic equator, as
nearly all of it has been displaced. This level of disk distortion is evident
regardless of where the observer is placed along the solar circle. This level of
disk warping is substantially beyond what is observed in the Galaxy, where the
offset between the Galactic plane and peak density of the stellar disk is at
most $1-2^\circ$ \citep{momany06} rather than the $\sim 10^\circ$ in the
simulation.

Despite this drawback, we find the simulation to be very useful in showing the
possible disk response morphologies generated by a substantial perturbation.
There are very clear ``streamers'' visible which appear to fly off from the disk
(in Figure~\ref{stelios_on} coincidentally overlapping where Feature~A is),
along with features of higher density and sharp edges up off the disk (near
Features~B and~C). As discussed above, we have intentionally selected the
position of the observer in this simulation to best highlight the agreement
between the simulation and the observations, so we should not over-interpret
this agreement as a conclusive statement about the origin of the MRi. The
presence of such strikingly similar features should lend credibility to the
hypothesis that the complex and highly structured MRi features in the Galaxy
have a common origin, but only if the degree of disk disruption can be brought
into line with observations.

Figure~\ref{stelios_combined} shows the nearer and further distance slices of
the \citet{kazantzidis09} model. The warp clearly extends across all of these
distance slices, with somewhat lower projected heights in the more distant
slice. However, these alternate slices show similar substructure
as in the ``mid'' slice, with no morphologically distinct components becoming
visible as is predicted by the \citet{penarrubia05} model. On this basis 
the tidal stream model is more easily seen to be in conflict with the
observations, but the disrupted disk may similarly be too extended in radius. A
quantiative comparison of the radial extent of the simulated disruption with the
observed MRi would be necessary to confirm this. The radial profile of the
disrupted disk material may also depend strongly on the disk's initial profile,
adding another degree of freedom to such models.

Based on these comparisons, the crucial question for future simulations to test
is whether such MRi-like features can be created without causing such an
unrealistically large distortion of the disk. This could be a matter of the
particular accretion history of the simulated disk, and less massive satellites
or particular infall trajectories could be more favorable. Including cold gas in
simulated accretion events could affect the outcome, possibly absorbing energy
of the infall or forming new stars within a distorted gas disk \citep[as in,
e.g., NGC 4565,][]{radburnsmith14}. The behavior of the gas may also provide
additional points of comparison between observations and simulations, as the
warping of the HI disk is well-studied \citep{kalberla09}.

Additionally, a more comprehensive search of the available parameter space in
the simulations could produce more realistic results. For example, the
\citet{kazantzidis09} simulations were run until the disk had ``settled'', in
that its bulk properties ceased to change significantly with time. While
reasonable for understanding the properties of disks under bombardment in
general, it is possible that we see the MRi today at a unique time in its
dynamical evolution, and a search of multiple timesteps in a simulation may show
transient effects (or possibly more evolved and settled states) that more
closely resemble the MRi.

\section{Discussion and Conclusions}
\label{sec_conclusions}

Exploiting the photometric accuracy and $3\pi$ sky coverage of the PS1 photometry,
we have presented the first distance-sensitive MSTO map of the Milky Way's stellar
distribution at low ($|b|<30^\circ$) Galactic latitudes. This map shows rich
substructure, much of which has been referred to as the MRi structure in the
past. This map (Figure~\ref{img_on}) presents the most complete and only
contiguous map of the MRi structure to date, showing its extent on both sides of
the Galactic plane and covering over $130^\circ$ in longitude. The
characteristic sharp edge in density at large heights above the disk is readily
distinguished on the northern and southern sides of the disk. The other arc-like
features in the north have positions and morphologies that are suggestive of a
connection to the MRi. The PS1 maps also suggest that the MRi is relatively
confined in radius, and we find no evidence of new structures related to the MRi
either closer to or further from the Sun.

It is not obvious how to decompose the structure we see in the MRi into a set of
distinct components. If the structure is the remnant of a tidal disruption, its position
along the Galactic disk certainly makes it challenging to recognize it as such.
As we have discussed, the superposition of a stream atop the exponential density
distribution of the disk could mask the signature of a stream. However, it is
difficult to see how an orbit (or orbits) could yield a distribution of material
over a narrow range in distance, since an accreted satellite must have fallen in
from large distances. This observed behavior in distance is a strong constraint
which future attempts to model an accretion event must agree with.

Likewise, it is difficult to intuitively understand what features are generated
when a stellar disk is disrupted by satellites. It is likely that the state of
the disk after such events is highly dependent on the mass and orbital
parameters of the satellite (or satellites), and additionally is likely to be
highly time-dependent. These difficulties in visualizing and modeling such
events should not cause us to exclude them.

The challenges of understanding and recreating either formation scenario
necessitate the use of models to both guide our understanding of the features we
see and to provide predictions that can be used for differentiating between
theories of the formation of such substructure. In both our comparisons to the
accretion model of \citet{penarrubia05} and perturbed disk model of
\citet{kazantzidis09} we find qualitative agreement in reproducing some of the
features of the MRi, but both also show areas of conflict with the observations.
Our objective with the PS1 maps is to show where these models need improvement,
and to show how even qualitative morphological constraints can be used to
further refine the models of different formation scenarios.

Though we have explored two models which are particularly well-suited for
comparison to the PS1 maps this is certainly not an exhaustive list of possible
models for the MRi. There are also analytical models for the MRi that we have
not considered in depth, such models that parameterize the structure as part of
a Galactic flare \citep{momany06,hammersley11}. Proper consideration of these
models requires a comprehensive fitting of the spatial and distance dependence
of the observed distribution of disk stars as a whole, which is beyond the scope
of this work. However, there are general morphological features of the flare
models that we can compare to the observations. The sharp edge in latitude that
has been characteristic of the MRi since its initial discovery, and which we
have shown also exists prominently in the southern hemisphere, is a particularly
strong constraint on Galactic flare models. Such a feature strongly suggests the
existence of some dynamically cold component with a low velocity dispersion,
which is at odds with flare models that require large vertical velocity
dispersions to raise stars to greater heights above the disk. 

Our presentation of these models is designed to link physical processes with the
morphological features they create on the sky, demonstrating where these
simulations perform best and drawing attention to where they most need
refinement in order to plausibly explain the observed substructure. In the case
of a tidal stream, we have shown that the challenge for future models is to
limit the debris to a compact range of Galactocentric radii while still filling
that range with a substantial amount of substructure. For perturbed disk models
the goal must be to create substantial structures out of the plane without
causing the disk to warp to such an unsupportable degree. The ability or
inability of future models to accommodate these conditions as dictated by the
data should help to narrow in on the origin of the MRi. This map provides an
obvious starting point for follow-up observations, as 3D velocities and
metallicities of the stars in the various ``features'' should help to untangle
their nature. Most of the stars should be bright enough to get good proper
motion estimates from {\it Gaia}, but radial velocities and metallicities may
require spectroscopy beyond the current set of spectroscopic surveys (RAVE,
SEGUE, APOGEE, LAMOST or {\it Gaia}). Nonetheless, it is clear that the Galactic
disk offers a rich example of how galaxy disks are being disturbed, and how they
respond to such disturbances.

\acknowledgments

We thank S. Kazantzidis for generously providing his simulation outputs.
CTS and EFB were supported during this work by NSF grant AST 1008342. EFS, EM,
and NFM acknowledge support from the DFG's SFB 881 grant ``The Milky Way
System'' (sub-project A3). NFM gratefully acknowledges the CNRS for support
through PICS project PICS06183. HWR acknowledges support from the European
Research Council under the European Union's Seventh Framework Programme (FP 7)
ERC Grant Agreement n. [321035].

The Pan-STARRS1 Survey has been made possible through contributions of the
Institute for Astronomy, the University of Hawaii, the Pan-STARRS Project
Office, the Max-Planck Society and its participating institutes, the Max Planck
Institute for Astronomy, Heidelberg and the Max Planck Institute for
Extraterrestrial Physics, Garching, The Johns Hopkins University, Durham
University, the University of Edinburgh, Queen’s University Belfast, the
Harvard-Smithsonian Center for Astrophysics, and the Las Cumbres Observatory
Global Telescope Network, Incorporated, the National Central University of
Taiwan, and the National Aeronautics and Space Administration under Grant No.
NNX08AR22G issued through the Planetary Science Division of the NASA Science
Mission Directorate.

{\it Facilities:} \facility{Pan-STARRS1}.

\end{document}